# Microwave Irradiation Assisted Deposition of $Ga_2O_3$ on III-nitrides for deep-UV opto-electronics

Piyush Jaiswal,[1,a)] Usman Ul Muazzam,[1] Anamika Singh Pratiyush,[1] Nagaboopathy Mohan,[1] Srinivasan Raghavan,[1] R. Muralidharan,[1] S. A. Shivashankar,[1] and Digbijoy N. Nath,[1]

[1] Centre for Nano Science and Engineering, Indian Institute of Science, 560012, Bangalore, India

We report on the deposition of $Ga_2O_3$ using microwave irradiation technique on III-nitride epi-layers. We also report on the first demonstration of a $Ga_2O_3$ device a visible blind deep -UV detector -with GaN-based heterostructure as the substrate. The film deposited in the solution medium, at $< 200^\circ C$, using a metalorganic precursor, was nanocrystalline. XRD confirms that as deposited film when annealed at high temperature turns polycrystalline $\beta-Ga_2O_3$. SEM shows the as-deposited film to be uniform, with a surface roughness of 4-5 nm, as revealed by AFM. Interdigitated metal semiconductor metal (MSM) devices with Ni/Au contact exhibited peak spectral response at 230 nm and a good visible rejection ratio. This first demonstration of a deep-UV detector on $\beta - Ga_2O_3$/III-nitride stack is expected to open up new possibilities of functional and physical integration of $\beta - Ga_2O_3$ and GaN material families towards enabling next generation high-performance devices by exciting band and heterostructure engineering.

## I. Introduction

Gallium oxide ($Ga_2O_3$) has 're-emerged' recently as a technologically important material, particularly for its promises in the areas of high-power electronics and deep-UV opto-electronics. In the past, it had been investigated for different applications such as gas sensing[1], phosphors[2], transparent electronic devices[3], spintronic devices[4], and in the field of catalysis[5]. In their seminal paper Roy et al.[6] described various polymorphs of $Ga_2O_3$: α-, β-, γ-, δ- and ε-$Ga_2O_3$ phases, which are relevant to bulk crystals as well as to micro- and nano- crystalline forms. Of these, β-$Ga_2O_3$, thermodynamically the most stable[12], is a wide band gap (4.9-5.3 eV)[7] semiconductor with a dielectric constant of 10.2−14.2 and breakdown field strength ($E_{br}$) of 8 $MV/cm^8$, having the added advantage of controllable electron densities[14] between $10^{15}$-$10^{19}$ $cm^{-3}$ and Baliga's FOM[1] ($\mu\varepsilon E_{br}^3$) of 2000-3400[9], which makes $Ga_2O_3$-based field effect transistors (FETs) promising for high power switching applications. Besides, among the various wide band gap materials which are naturally attractive for solar-blind deep-UV detection, $Ga_2O_3$ doesn't suffer from material challenges, such as misfit dislocations and stacking faults, which plague the rival III-nitride system[10] due to the requirement of growth on foreign substrates. While phase segregation between ZnO wurtzite and MgO rock salt in MgO-rich ZnMgO[11], and too large a band gap for diamond, are disadvantages of other wide band gap systems, $Ga_2O_3$ being a binary compound with an absorption edge at 240-250 nm[12-15], suffers from none of these drawbacks. One of the most critical advantages it enjoys is the prospect of scalability and economic viability, as large-area single crystal boules of $Ga_2O_3$ can be grown using bulk growth techniques such as float-zone[16] and Czochralski[17], which opens the possibility of having inexpensive, high-quality, large-area single crystal wafers as substrates.

While $Ga_2O_3$ grown/deposited on bulk substrates[18] or on sapphire[11,13] are widely reported, studies on $Ga_2O_3$-based heterostructures are still at an embryonic stage. In particular, integration of $Ga_2O_3$ with dissimilar materials could open up new vistas for exciting heterostructure and band engineering. Growth of $Ga_2O_3$ on III-nitrides, for instance, could be promising for both physical and functional integration in terms of leveraging the advantages of emerging $Ga_2O_3$ devices with the benefits of a well-established and mature GaN-based technology. By exploiting the band alignments and band gap tailoring of the $Ga_2O_3$-GaN material families, novel and functional devices could be realized, besides exploring the physics of carrier transport across such junctions.

In this work, we report on the deposition of $Ga_2O_3$ on III-nitride layers in the solution medium using a microwave irradiation technique. We also report the first demonstration of a $Ga_2O_3$ device – a solar-blind deep-UV detector - on GaN-based epi-layers as the substrate.

a) Corresponding author email: piyushj@iisc.ac.in

Currently, $Ga_2O_3$ thin films for solar-blind PD application are being fabricated/deposited by techniques such as cation exchange between layers[19], molecular beam epitaxy (MBE)[14,15,20], surface oxidation[21], chemical vapour deposition and related techniques (MOCVD)[22], and other physical vapor deposition techniques[23]. All these are generally expensive and time-consuming and require high substrate high temperatures. Here we report on $Ga_2O_3$ layer formation on epitaxial GaN, through microwave irradiation of a reactant solution. Microwave irradiation-based deposition in the solution medium has been shown to provide oxide films of high quality[24]. Here we have obtained $Ga_2O_3$ films for device application through the simple and rapid microwave irradiation-assisted deposition process using commercial (standard) apparatus[25] (see supplementary material for details of deposition process). Epitaxially grown GaN layers (figure 4(a), see Raghavan et. al.[26] for HEMT stack growth details) were used as substrates, because of the close lattice match between h-GaN and $\beta$-$Ga_2O_3$ along the (100) direction. In addition, GaN is suitable for handling high voltages and current densities, and for operation at high switching speeds.

Figure 1 shows the XPS analysis to study the chemical composition and bonding states of the deposited thin films. The absence of any $N_2$ photopeak shows the uniform coverage of film over the AlGaN/GaN layer. The survey spectra (Figure 2a) shows relevant photoelectron peaks for gallium (Ga 2p1/2, Ga 2p3/2, Ga 3s, Ga 3p, and Ga 3d), oxygen (O 1s, O 2s) and carbon (C 1s), together with the Auger lines from gallium (GaLMM) and oxygen (OKVV, OKLL)[27-29]. The presence of C 1s photopeak indicates the presence of adventitious carbon due to the organic solvents used during post deposition cleaning of the film. The binding energies of the Ga $2p_{3/2}$ (1118.1 eV), Ga $3d_{5/2}$ (20.5 eV) and O 1s (531.2 eV) (figure 1b, c and d respectively) photoelectron peaks obtained from a high-resolution spectrum indicate that the gallium is in the 3+ oxidation state $(Ga^{3+})$[30]. The peak at 24.8 eV (figure 1d) corresponds to O 2s core level[31] and at 529.9 eV (figure 1c) suggests $O_2$ vacancy[32]. The Ga/O ratio (1.01) confirms this $O_2$ deficiency and a n-type film, which is not surprising as most of the semiconducting oxides and low temperature deposited films are naturally n-type including $\beta$-$Ga_2O_3$[29,33]. Comparing these observations with those reported in literature[28,29,34] it can be concluded that the deposited film is gallium oxide (albeit gallium rich n-type $Ga_2O_{1.97}$).

Figure 2 shows the XRD (Rigaku SmartLab©) patterns of the bare substrate (h-GaN), the as-deposited oxide film on it, and the oxide film rapid thermal-annealed for 10 min in oxygen ambient at 950°C [Annealsys© *As-One* RTP system], in blue, red, and green colour, respectively. The absence of any substrate (GaN) peak in the pattern of the as-deposited film implies that the oxide film is sufficiently thick (70-80 nm, figure 3b). The broad peaks testify that the film is nano-crystalline (3.3 nm avg. crystallite size, by the Scherrer formula). The peaks can nevertheless be indexed to the $\gamma$-phase of $Ga_2O_3$[35], which transform to $\beta$-phase[6] as confirmed by the XRD pattern of the well-crystallised rapid-annealed film (22.4 nm – avg. crystallite size)[36]. The counter-intuitive formation of $\gamma$-$Ga_2O_3$ at sub-200°C is attributable to locally elevated temperatures in the irradiated solution and to the nucleation kinetics of $\gamma$-$Ga_2O_3$[37] formation because $\gamma$-$Ga_2O_3$ has a spinel structure associated with many vacancies[38] and crystals containing vacant sites are stabilized at low crystallization temperatures.

Substantiating the XRD and XPS results, Figure 3(a) shows the SEM (GEMINI Ultra 55 FE-SEM, Carl Zeiss©) of the as-deposited film uniformly coating the entire substrate; the corresponding EDS is consistent with a gallium-rich oxide layer. The SEM data show also that the film is nano-crystalline (figure 3c) and its thickness is ~73 nm (figure 3b). AFM data confirms that the crystallite size is substantially larger in the annealed film (figure 3d). As may be expected, annealing enhances the roughness (15 nm, figure 3f) of the annealed film vis-à-vis the as-deposited film (~5 nm, figure 3e).

Following standard lithographic procedures using i-line optical lithography, Ni/Au (20/70 nm) was e-beam-evaporated to make Schottky contacts to the $Ga_2O_3$/GaN/AlGaN HEMT stack, as shown in figure 4 (a). Figure 4(b) shows optical micrograph of the metal-semiconductor-metal (MSM) photodetector (PD) fabricated, comprising 16 pairs of inter-digitated fingers 5 μm wide, spacing of 6 μm, and an active area of 250 x 300 μm². For spectral responsivity measurements, a quantum efficiency (QE) set-up from Sciencetech© Inc.) was used. The QE set-up consists of a xenon lamp (150 W), monochromator, an optical chopper (25 Hz), a light-focusing lens assembly, a lock-in amplifier and a stage for electrical probing of the PD. A standard UV-enhanced Si photodiode was used for the calibrating the lamp power of the QE set-up, prior to the measurements. White light illumination from the xenon lamp was chopped at 25 Hz and focused onto the PD after being guided through the monochromator and the focusing lens assembly. The the photocurrent generated from the PD was amplified and measured using a lock-in amplifier (Stanford SR810), whose spectral responsivity (SR) output was obtained in

a) Corresponding author email: piyushj@iisc.ac.in

Amperes/Watt. The current-voltage characteristics (photo- and dark current) of the PD was measured using sourcemeter (Keithley 2450) externally connected to the QE set-up.

Figure 4(c) shows variation of the SR with wavelength as a function of applied voltage on the log scale (22 V, 25 V, 30 V, and 35 V). The inset to figure 5 shows variation of SR with wavelength on a linear scale. The peak SR was found to be at 236 nm, consistent with reports in the literature[39]. Peak responsivity values of 0.1 A/W and 0.8 A/W were measured at the bias voltages of 22 and 35 V, respectively. The UV-to-visible rejection ratio was calculated by dividing the SR value at 230 nm (peak SR) by that at 400 nm. A rejection ratio >$10^3$ was observed in the visible at a bias of 22 V, testifying to the visible-blind nature of the PD. A kink was also observed at 365 nm. This second peak can be attributed to the absorption of photons corresponding to 365 nm (~3.4 eV) by the epitaxial GaN layer underneath the $Ga_2O_3$ film.

Figure 4(d) shows current-voltage (I-V) characteristics of MSM-based $Ga_2O_3$ PD. With an applied voltage of 20 V, the measured dark current was found to be ~12 nA, which is attributed to the highly resistive $Ga_2O_3$ film. However, as the illumination (230 nm) was turned on, the photocurrent was found to be ~ 82 nA at 20 V. There was an increase in current by one order of magnitude upon UV illumination.

Figure 4(e) shows the variation of peak SR at an illumination of 230 nm with applied bias voltages from 22 V to 35 V. The peak SR was found to increase with increase in bias voltage, as shown in figure 6. This indicates the presence of an internal gain.[15,40,41]

With an illumination wavelength of 230 nm (with a power of 11 $mAW/cm^2$), and an applied bias of 22 V, the peak SR was found to be 0.1 A/W (as shown in figure 4(c)). As the bias voltage was increased to 35 V, the peak SR increased to 0.8 A/W. By assuming the external quantum efficiency (η) to be 100%, the theoretical responsivity ($R_{Th}$) of UV photodetectors having a detection range of 230 nm can be calculated using the expression below:

$$R_{Th} = \frac{\eta q}{h\nu} \qquad \qquad 1(a)$$

$$R_{Calculated} = \frac{G\eta q}{h\nu} \qquad \qquad 1(b)$$

where G is gain in the PD, hν is the energy. The theoretical SR value for 230 nm comes out to be 0.18 A/W. This value of ideal responsivity is surpassed above the bias voltage of 25 V, which indicates gain in the PD[40,41]. This gain comes from either an oxygen-deficient film, leading to trapping of holes in bulk, or from interface states at the metal-semiconductor (M-S) Schottky junction enabling hole trapping at the M-S junction itself[15,41,42]. As the M-S junction has to maintain charge neutrality, more electrons have to flow from the metal side, subsequently lowering the Schottky barrier, thereby leading to gain in the PD. A similar phenomenon has been reported in GaN, AlGaN, and β-$Ga_2O_3$ MSM-based PDs earlier[15,41-43].

In conclusion, we reported on the microwave irradiation assisted deposition of β−$Ga_2O_3$ on III-nitride epi-layers as the substrate. The thin films of β − $Ga_2O_3$ were polycrystalline in nature and were characterized by SEM, AFM, XRD and XPS. A visible blind UV A visible blind UV detector with good spectral response and photocurrent was demonstrated on the β-$Ga_2O_3$ deposited on AlGaN/GaN HEMT. This first demonstration of a device on β-$Ga_2O_3$/III-nitride stack is expected to open up new possibilities of functional and physical integration of β-$Ga_2O_3$ and GaN material families towards enabling next generation high-performance devices.

## II. Supplementary material

See supplementary material for details of the microwave irradiation assisted $Ga_2O_3$ deposition technique.

## III. Acknowledgement

This work is funded by DST WTI (grant. No. 01519) and the ISRO-IISc Space Technology Cell (STC).

[a] Corresponding author email: piyushj@iisc.ac.in

a) Corresponding author email: piyushj@iisc.ac.in

[a] Corresponding author email: piyushj@iisc.ac.in

**Figures and legends**

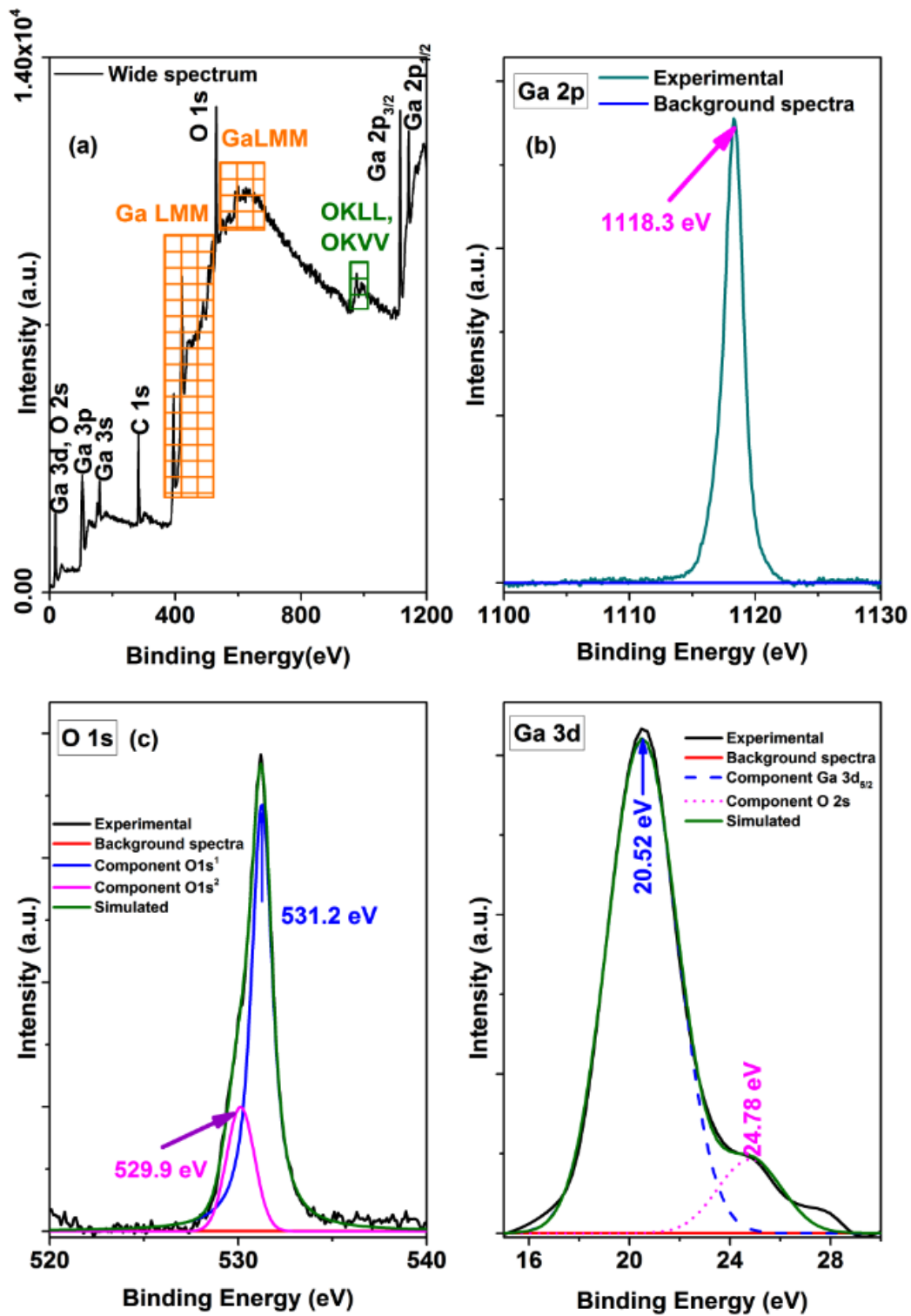

**Figure 1**: (a) Survey spectra of the as deposited film showing all the expected peaks of $Ga_2O_3$ (b) $Ga2p_{3/2}$ peaks at 1118.3 eV signifies typical Ga-O bond value in $Ga_2O_3$ (c) O 1s peak at 531.2 eV is the expected BE for Ga-O bond and the one at 529.9 eV indicates $O_2$ vacancy (d) $Ga3d_{5/2}$ peak at 20.52 eV is denotes the d-shell Ga-O BE and the one at 24.7 eV arises due to O 2s core level

a) Corresponding author email: piyushj@iisc.ac.in

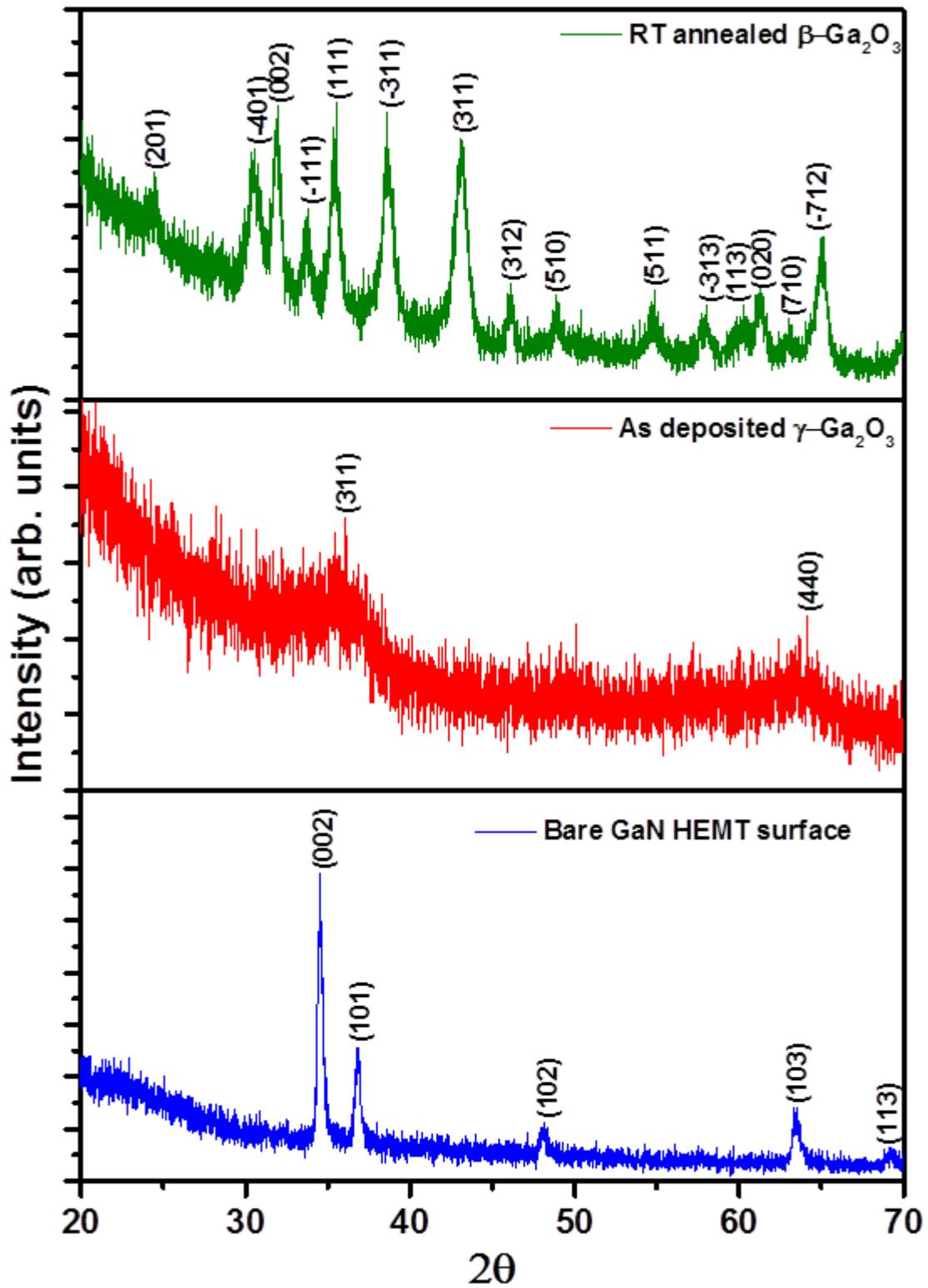

**Figure 2**: XRD diffraction patterns of bare substrate – hexagonal GaN (blue), as deposited (red) nano-crystalline β-$Ga_2O_3$ over h-GaN and the same rapid thermal (RT) annealed (green) to polycrystalline β-$Ga_2O_3$ at 950ºC for 10 min under 100 sccm $O_2$

[a] Corresponding author email: piyushj@iisc.ac.in

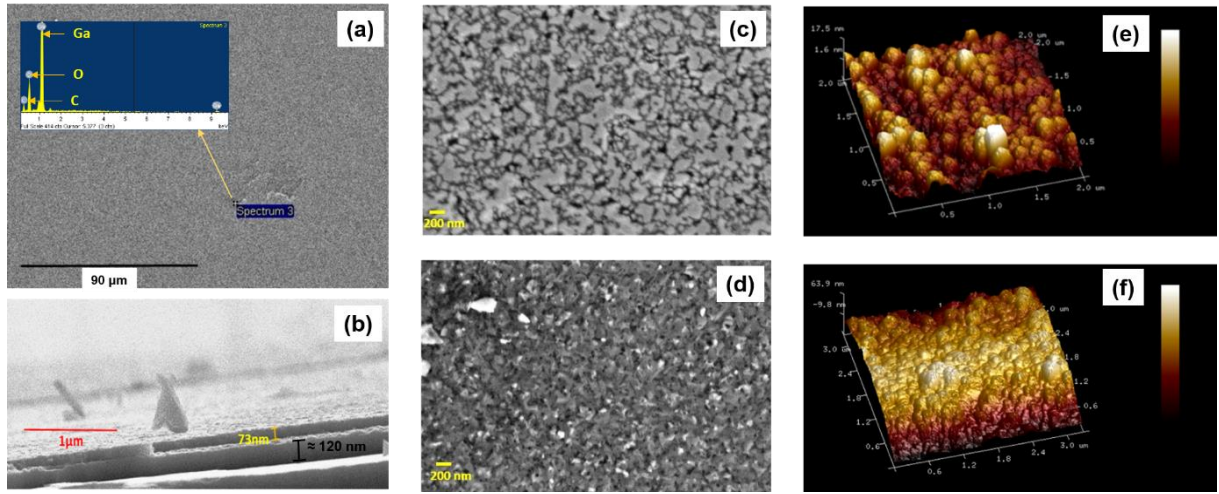

**Figure 3**: (a) SEM micrograph showing uniform surface coverage by as deposited $Ga_2O_3$ film with the corresponding EDS (inset) data revealing a gallium rich layer (b) the cross-sectional even thickness of 73 nm (c) showing the film is made up of nano-crystallites (3.3 nm avg. size) (d) film turning to polycrystalline when rapidly heated at 950ºC with avg. grain size of 22 nm (e) as deposited smooth film with $R_{rms}$ of 4-5 nm (f) showing increasing $R_{rms}$ to 10-15 nm when annealed

[a] Corresponding author email: piyushj@iisc.ac.in

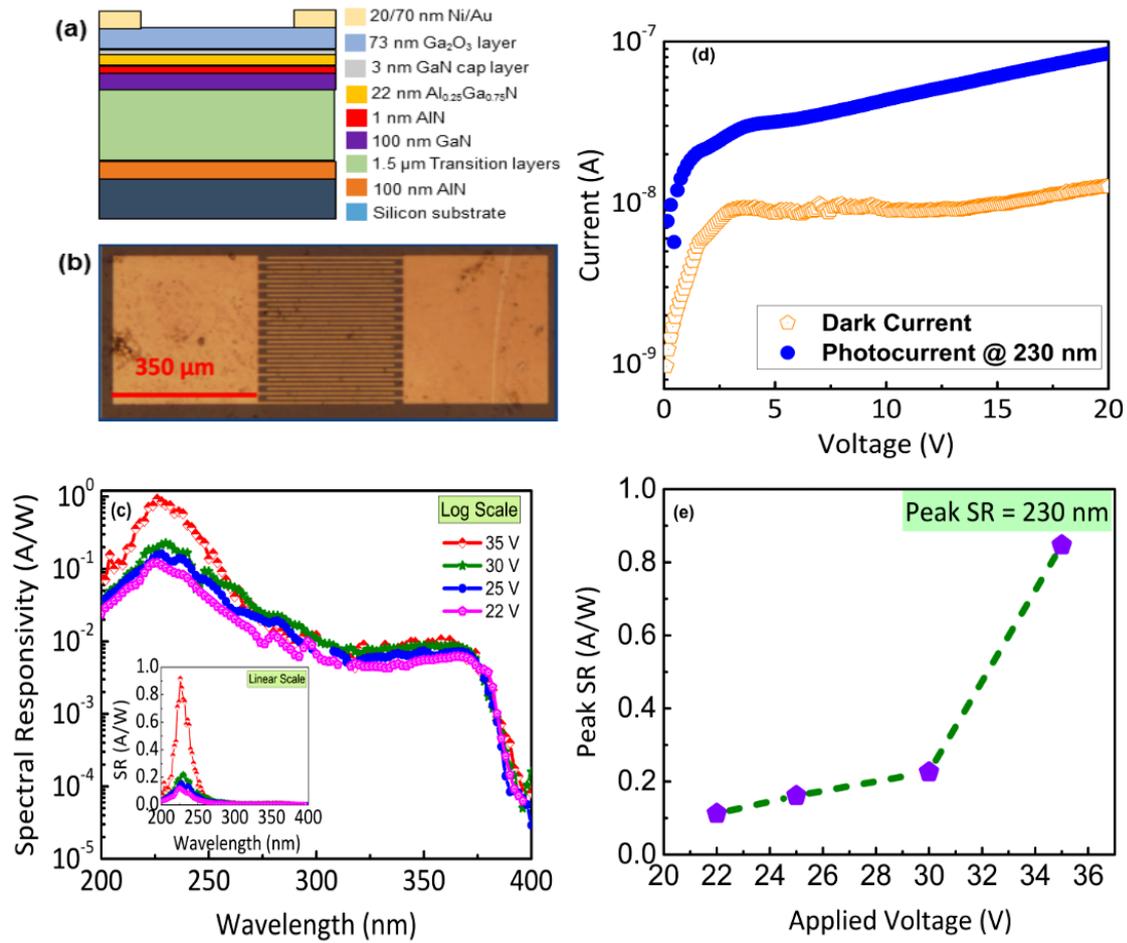

**Figure 4**: (a) Schematic of AlGaN/GaN HEMT stack (Side-view)[26] (b) Optical micrograph of the fabricated MSM device (Top-view).(c) Variation of spectral response with wavelength in log scale as a function of bias voltages (22 V to 35 V), also showing UV to visible rejection ratio >$10^3$ at 22 V. The inset shows variation of SR with wavelength as a function of bias voltages in linear scale (d) Variation of dark, and photocurrent with applied bias. The photo current was measured under 230 nm illumination (e) Variation of peak responsivity (SR at 230 nm) with applied bias, (measured at an optical chopping frequency of 30 Hz).

a) Corresponding author email: piyushj@iisc.ac.in

# Supplementary material: Microwave Irradiation Assisted Deposition of Ga2O3 on III-nitrides for deep-UV opto-electronics

Piyush Jaiswal,[*] Usman Ul Muazzam, Anamika Singh Pratiyush, Nagabhoopathy Mohan, Srinivasan Raghavan, R. Muralidharan, S.A.Shivashankar, and Digbijoy N Nath

Centre for Nano Science and Engineering, Indian Institute of Science, Bangalore 560012, India
(Dated: October 25, 2017)

**Process details: Microwave irradiation assisted deposition of Ga₂O₃ thin film on GaN based HEMT substrate**

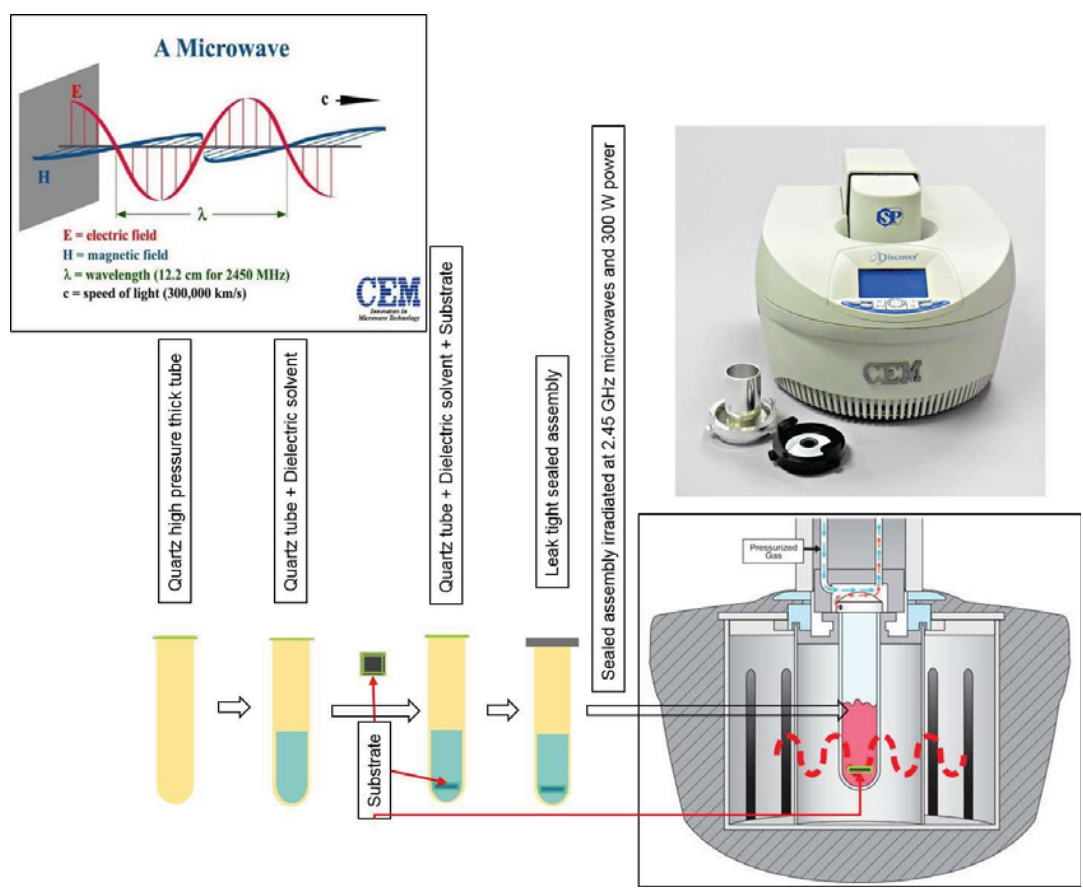

FIG. 1: Schematic representation of the step-wise microwave irradiation assisted film deposition process

β-diketonate complexes of Ga, namely, Ga(III) acetylacetonate (Ga(acac)₃), was used in the synthesis. AR-grade ethanol, 1-decanol were utilized as solvents. Solutions of 1 mmol of Ga(acac)₃ in 40 ml of alcoholic solution (1-decanol:ethanol::5:3) were transferred to the reaction vessel. A GaN based HEMT stack piece (1 cm x 1 cm) was cleaned by standard solvent cleaning protocol and immersed in the solution in the vessel. This reaction mixture was irradiated in a single-mode microwave reactor (© CEM Corp., USA, 2.45 GHz, 300 W), shown in Fig. 1, for 10 min, yielding a uniform coating on the substrate as well as a small amount of precipitate at the bottom of the vessel.

*Corresponding author: piyushj@iisc.ac.in

The solid precipitate was separated by centrifugation and recovered, whereas the coated GaN HEMT substrate was cleaned by ultrasonication in acetone and ethanol for 5 minutes.

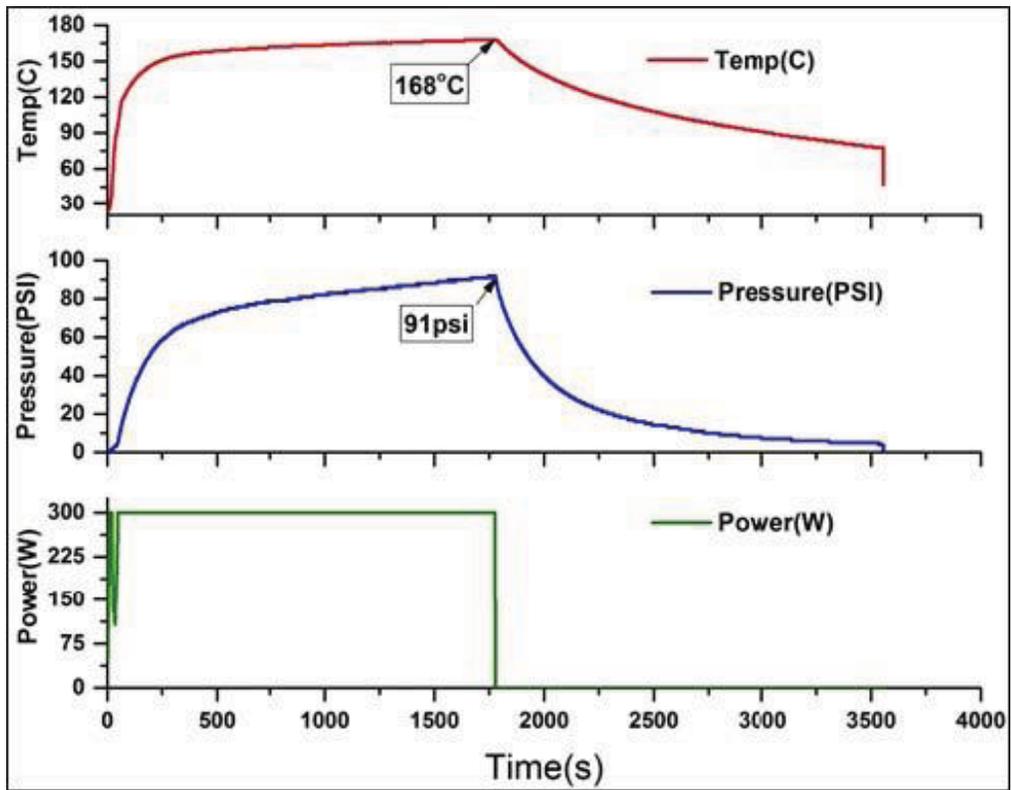

FIG. 2: Deposition parameters during the microwave irradiation assisted film deposition process

The temperature of the reaction mixture was < 200$^{\circ}$C during microwave irradiation with pressures less than 100 psi during the process under a constant field microwave power of 300 W (Fig. 2). The ramping up (20 min), the microwave irradiation (10 min) and the cooling time (till 60$^{\circ}$C) (25 min) took less than an hour for the whole deposition process to complete.

*Corresponding author: piyushj@iisc.ac.in